\documentclass[%
 reprint,
 amsmath,amssymb,superscriptaddress,prl
]{revtex4-1}
\usepackage{hyperref}
\hypersetup{colorlinks=true, linkcolor=blue, urlcolor=blue, citecolor=blue}
\usepackage{cleveref}
\usepackage{graphicx}
\usepackage{epstopdf}
\usepackage{dcolumn}
\usepackage{bm}
\usepackage{breqn}
\usepackage{xspace}

\makeatletter
\let\cat@comma@active\@empty
\makeatother

\begin{document}


\title{Searches for Sterile Neutrinos with the IceCube Detector}

\affiliation{III. Physikalisches Institut, RWTH Aachen University, D-52056 Aachen, Germany}
\affiliation{Department of Physics, University of Adelaide, Adelaide, 5005, Australia}
\affiliation{Dept.~of Physics and Astronomy, University of Alaska Anchorage, 3211 Providence Dr., Anchorage, AK 99508, USA}
\affiliation{CTSPS, Clark-Atlanta University, Atlanta, GA 30314, USA}
\affiliation{School of Physics and Center for Relativistic Astrophysics, Georgia Institute of Technology, Atlanta, GA 30332, USA}
\affiliation{Dept.~of Physics, Southern University, Baton Rouge, LA 70813, USA}
\affiliation{Dept.~of Physics, University of California, Berkeley, CA 94720, USA}
\affiliation{Lawrence Berkeley National Laboratory, Berkeley, CA 94720, USA}
\affiliation{Institut f\"ur Physik, Humboldt-Universit\"at zu Berlin, D-12489 Berlin, Germany}
\affiliation{Fakult\"at f\"ur Physik \& Astronomie, Ruhr-Universit\"at Bochum, D-44780 Bochum, Germany}
\affiliation{Physikalisches Institut, Universit\"at Bonn, Nussallee 12, D-53115 Bonn, Germany}
\affiliation{Universit\'e Libre de Bruxelles, Science Faculty CP230, B-1050 Brussels, Belgium}
\affiliation{Vrije Universiteit Brussel, Dienst ELEM, B-1050 Brussels, Belgium}
\affiliation{Dept.~of Physics, Massachusetts Institute of Technology, Cambridge, MA 02139, USA}
\affiliation{Dept.~of Physics, Chiba University, Chiba 263-8522, Japan}
\affiliation{Dept.~of Physics and Astronomy, University of Canterbury, Private Bag 4800, Christchurch, New Zealand}
\affiliation{Dept.~of Physics, University of Maryland, College Park, MD 20742, USA}
\affiliation{Dept.~of Physics and Center for Cosmology and Astro-Particle Physics, Ohio State University, Columbus, OH 43210, USA}
\affiliation{Dept.~of Astronomy, Ohio State University, Columbus, OH 43210, USA}
\affiliation{Niels Bohr Institute, University of Copenhagen, DK-2100 Copenhagen, Denmark}
\affiliation{Dept.~of Physics, TU Dortmund University, D-44221 Dortmund, Germany}
\affiliation{Dept.~of Physics and Astronomy, Michigan State University, East Lansing, MI 48824, USA}
\affiliation{Dept.~of Physics, University of Alberta, Edmonton, Alberta, Canada T6G 2E1}
\affiliation{Erlangen Centre for Astroparticle Physics, Friedrich-Alexander-Universit\"at Erlangen-N\"urnberg, D-91058 Erlangen, Germany}
\affiliation{D\'epartement de physique nucl\'eaire et corpusculaire, Universit\'e de Gen\`eve, CH-1211 Gen\`eve, Switzerland}
\affiliation{Dept.~of Physics and Astronomy, University of Gent, B-9000 Gent, Belgium}
\affiliation{Dept.~of Physics and Astronomy, University of California, Irvine, CA 92697, USA}
\affiliation{Dept.~of Physics and Astronomy, University of Kansas, Lawrence, KS 66045, USA}
\affiliation{Dept.~of Astronomy, University of Wisconsin, Madison, WI 53706, USA}
\affiliation{Dept.~of Physics and Wisconsin IceCube Particle Astrophysics Center, University of Wisconsin, Madison, WI 53706, USA}
\affiliation{Institute of Physics, University of Mainz, Staudinger Weg 7, D-55099 Mainz, Germany}
\affiliation{Department of Physics, Marquette University, Milwaukee, WI, 53201, USA}
\affiliation{Universit\'e de Mons, 7000 Mons, Belgium}
\affiliation{National Research Nuclear University MEPhI (Moscow Engineering Physics Institute), Moscow, Russia}
\affiliation{Physik-department, Technische Universit\"at M\"unchen, D-85748 Garching, Germany}
\affiliation{Institut f\"ur Kernphysik, Westf\"alische Wilhelms-Universit\"at M\"unster, D-48149 M\"unster, Germany}
\affiliation{Bartol Research Institute and Dept.~of Physics and Astronomy, University of Delaware, Newark, DE 19716, USA}
\affiliation{Dept.~of Physics, Yale University, New Haven, CT 06520, USA}
\affiliation{Dept.~of Physics, University of Oxford, 1 Keble Road, Oxford OX1 3NP, UK}
\affiliation{Dept.~of Physics, Drexel University, 3141 Chestnut Street, Philadelphia, PA 19104, USA}
\affiliation{Physics Department, South Dakota School of Mines and Technology, Rapid City, SD 57701, USA}
\affiliation{Dept.~of Physics, University of Wisconsin, River Falls, WI 54022, USA}
\affiliation{Oskar Klein Centre and Dept.~of Physics, Stockholm University, SE-10691 Stockholm, Sweden}
\affiliation{Dept.~of Physics and Astronomy, Stony Brook University, Stony Brook, NY 11794-3800, USA}
\affiliation{Dept.~of Physics, Sungkyunkwan University, Suwon 440-746, Korea}
\affiliation{Dept.~of Physics, University of Toronto, Toronto, Ontario, Canada, M5S 1A7}
\affiliation{Dept.~of Physics and Astronomy, University of Alabama, Tuscaloosa, AL 35487, USA}
\affiliation{Dept.~of Astronomy and Astrophysics, Pennsylvania State University, University Park, PA 16802, USA}
\affiliation{Dept.~of Physics, Pennsylvania State University, University Park, PA 16802, USA}
\affiliation{Dept.~of Physics and Astronomy, University of Rochester, Rochester, NY 14627, USA}
\affiliation{Dept.~of Physics and Astronomy, Uppsala University, Box 516, S-75120 Uppsala, Sweden}
\affiliation{Dept.~of Physics, University of Wuppertal, D-42119 Wuppertal, Germany}
\affiliation{DESY, D-15735 Zeuthen, Germany}

\author{M.~G.~Aartsen}
\affiliation{Department of Physics, University of Adelaide, Adelaide, 5005, Australia}
\author{K.~Abraham}
\affiliation{Physik-department, Technische Universit\"at M\"unchen, D-85748 Garching, Germany}
\author{M.~Ackermann}
\affiliation{DESY, D-15735 Zeuthen, Germany}
\author{J.~Adams}
\affiliation{Dept.~of Physics and Astronomy, University of Canterbury, Private Bag 4800, Christchurch, New Zealand}
\author{J.~A.~Aguilar}
\affiliation{Universit\'e Libre de Bruxelles, Science Faculty CP230, B-1050 Brussels, Belgium}
\author{M.~Ahlers}
\affiliation{Dept.~of Physics and Wisconsin IceCube Particle Astrophysics Center, University of Wisconsin, Madison, WI 53706, USA}
\author{M.~Ahrens}
\affiliation{Oskar Klein Centre and Dept.~of Physics, Stockholm University, SE-10691 Stockholm, Sweden}
\author{D.~Altmann}
\affiliation{Erlangen Centre for Astroparticle Physics, Friedrich-Alexander-Universit\"at Erlangen-N\"urnberg, D-91058 Erlangen, Germany}
\author{K.~Andeen}
\affiliation{Department of Physics, Marquette University, Milwaukee, WI, 53201, USA}
\author{T.~Anderson}
\affiliation{Dept.~of Physics, Pennsylvania State University, University Park, PA 16802, USA}
% ...

\date{\today}

\collaboration{IceCube Collaboration}
\thanks{please direct correspondence to: analysis@icecube.wisc.edu}
\noaffiliation


\begin{abstract}
The IceCube neutrino telescope at the South Pole has measured the atmospheric muon neutrino spectrum as a function of zenith angle and energy in the approximate 320~GeV to 20~TeV range, to search for the oscillation signatures of light sterile neutrinos.  No evidence for anomalous $\nu_\mu$ or $\overline{\nu}_\mu$ disappearance is observed in either of two independently developed analyses, each using one year of atmospheric neutrino data. New exclusion limits are placed on the parameter space of the 3+1 model, in which muon antineutrinos would experience a strong Mikheyev-Smirnov-Wolfenstein-resonant oscillation. The exclusion limits extend to $\mathrm{sin}^2 2\theta_{24} \leq$  0.02 at $\Delta m^2 \sim$ 0.3 $\mathrm{eV}^2$ at the 90\% confidence level. The allowed region from global analysis of appearance experiments, including LSND and MiniBooNE, is excluded at approximately the 99\% confidence level for the global best fit value of $|$U$_{e4}|^2$.
\end{abstract}


\maketitle

\newpage
\clearpage

\section{\label{sec:AtmosNu}Introduction }
 Sterile neutrinos with masses in the range $\Delta m^2 = 0.1$~${\rm eV}^2 - 10~{\rm eV}^2$ have been posited to explain anomalies in accelerator \cite{Athanassopoulos:1996jb,Aguilar:2001ty,Aguilar-Arevalo:2013pmq}, reactor \cite{Mention:2011rk}, and radioactive source~\cite{Bahcall:1994bq}~oscillation experiments.  Several null results \cite{Armbruster:2002mp,Abe:2014gda, Adamson:2011ku, Cheng:2012yy, Dydak:1983zq} restrict the available parameter space of the minimal 3+1 model, which assumes mixing of the three active neutrinos with a single sterile neutrino, resulting in three light and one heavier mass state. Global fits to world data \cite{Giunti:2011gz,Kopp:2013vaa,Conrad:2012qt} demonstrate that there remain regions of allowed parameter space around the best fit point of $\Delta m^2 = 1~{\rm eV}^2$ and $\sin^2 2\theta_{24}=0.1$. A consequence of these models is the existence of $\nu_\mu$ ($\bar{\nu}_\mu$) disappearance signatures, which are yet to be observed.  

Atmospheric neutrinos produced in cosmic ray air showers throughout the Earth's atmosphere are detected by IceCube \cite{Halzen:2010yj}.  To mitigate the large atmospheric muon background, only up-going neutrinos are selected. For these trajectories, the Earth acts as a filter to remove the charged particle background. At high neutrino energies, the Earth also modifies the neutrino flux due to charged current and neutral current interactions \cite{GonzalezGarcia:2005xw}. At E$_\nu~>~$100~GeV, oscillations due to the known neutrino mass splittings have wavelengths larger than the diameter of the Earth and can be neglected.  

A previous measurement of the atmospheric flux in the sub-TeV range, performed by the Super-Kamiokande experiment, found no evidence for anomalous neutrino disappearance \cite{Abe:2014gda}. 
This paper reports the first searches for $(\nu_\mu+\overline{\nu}_\mu)$ disappearance in the approximate 320~GeV to 20~TeV range, using two independent analyses each based on one-year data samples from the IceCube detector \cite{Aartsen:2015rwa,Aartsen:2013eka}. In this energy regime, sterile neutrinos would produce distinctive energy-dependent distortions of the measured zenith angle distributions \cite{Nunokawa:2003ep}, caused by resonant matter-enhanced oscillations during neutrino propagation through the Earth.  

This MSW resonant effect depletes antineutrinos in 3+1 models (or neutrinos in 1+3) \cite{Nunokawa:2003ep,Abazajian:2012ys}.  Additional oscillation effects produced by sterile neutrinos include vacuum-like oscillations at low energies for both neutrinos and antineutrinos, and a modification of the Earth opacity  at high energies, as sterile neutrinos are unaffected by matter.  These effects would lead to detectable distortions of the flux in energy and angle, henceforth called ``shape effects,'' in IceCube for mass splittings in the range 0.01$~{\rm eV}^2\leq \Delta m^2 \leq~10~{\rm eV}^2$ \cite{Choubey:2007ji, Razzaque:2011ab, Barger:2011rc, Esmaili:2012nz, Razzaque:2012tp,Esmaili:2013vza, Esmaili:2013cja, Lindner:2015iaa}.

\begin{figure}[tbp!]
\includegraphics[width=1.0\columnwidth]{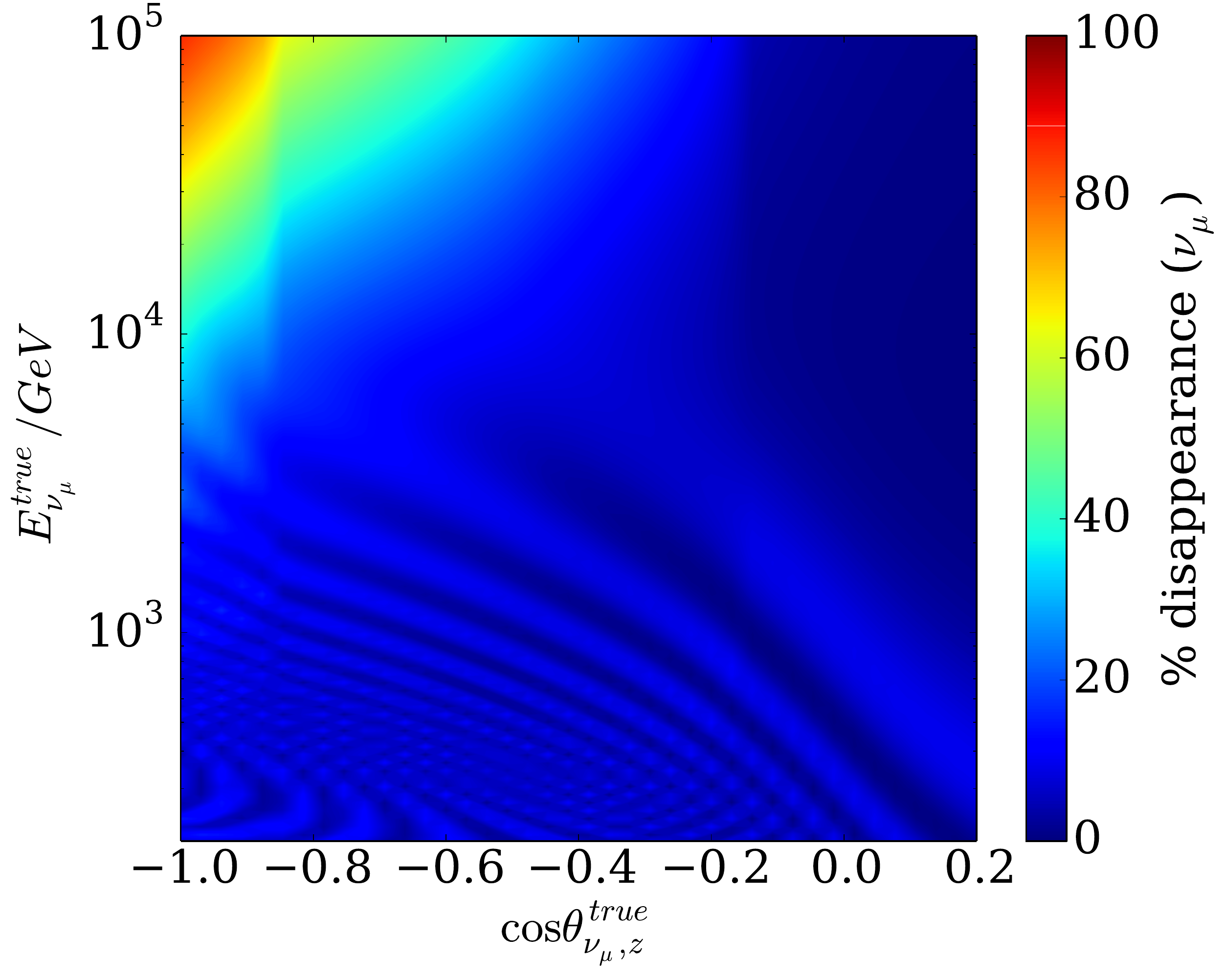}\\
\includegraphics[width=1.0\columnwidth]{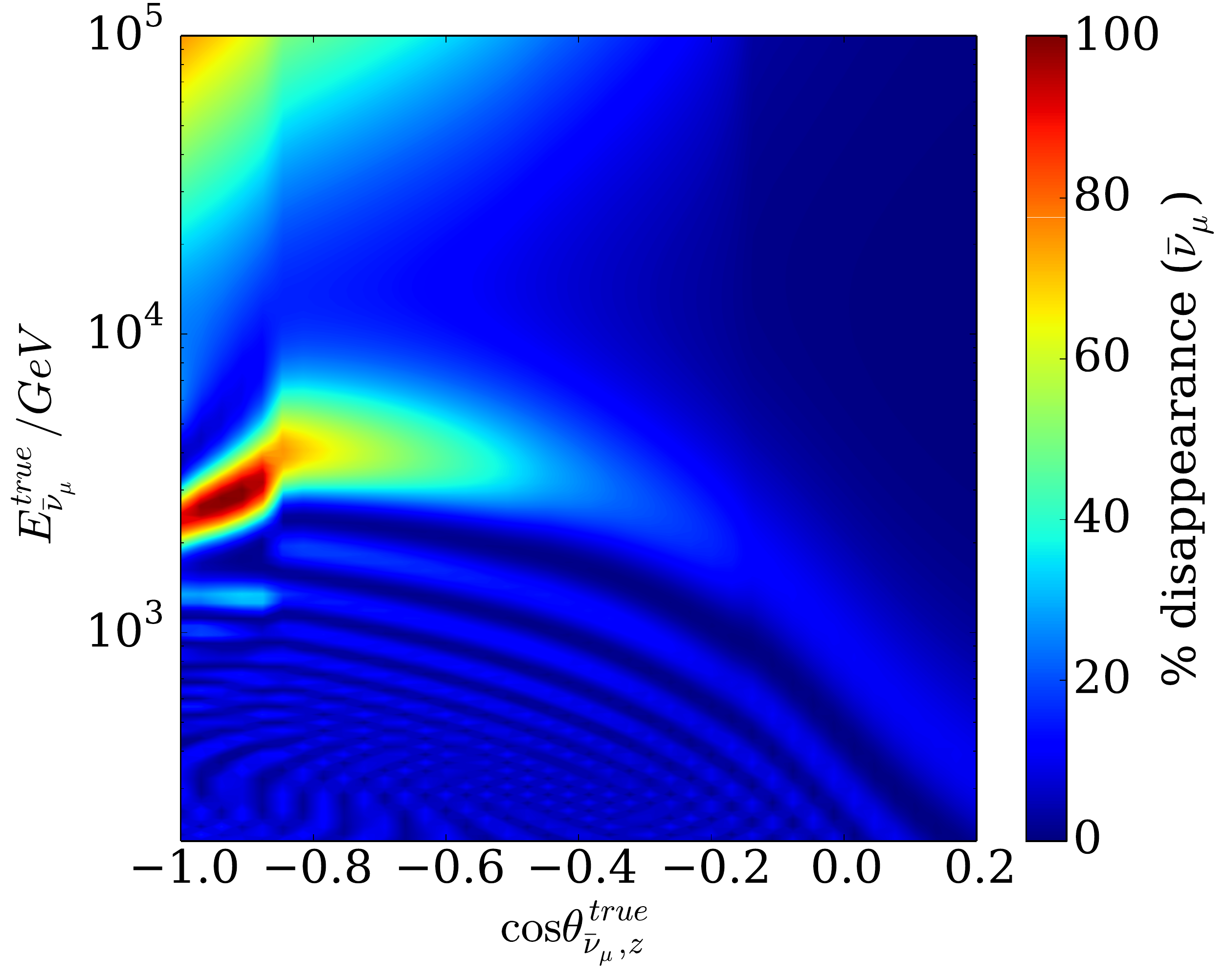}\\
\includegraphics[width=1.0\columnwidth]{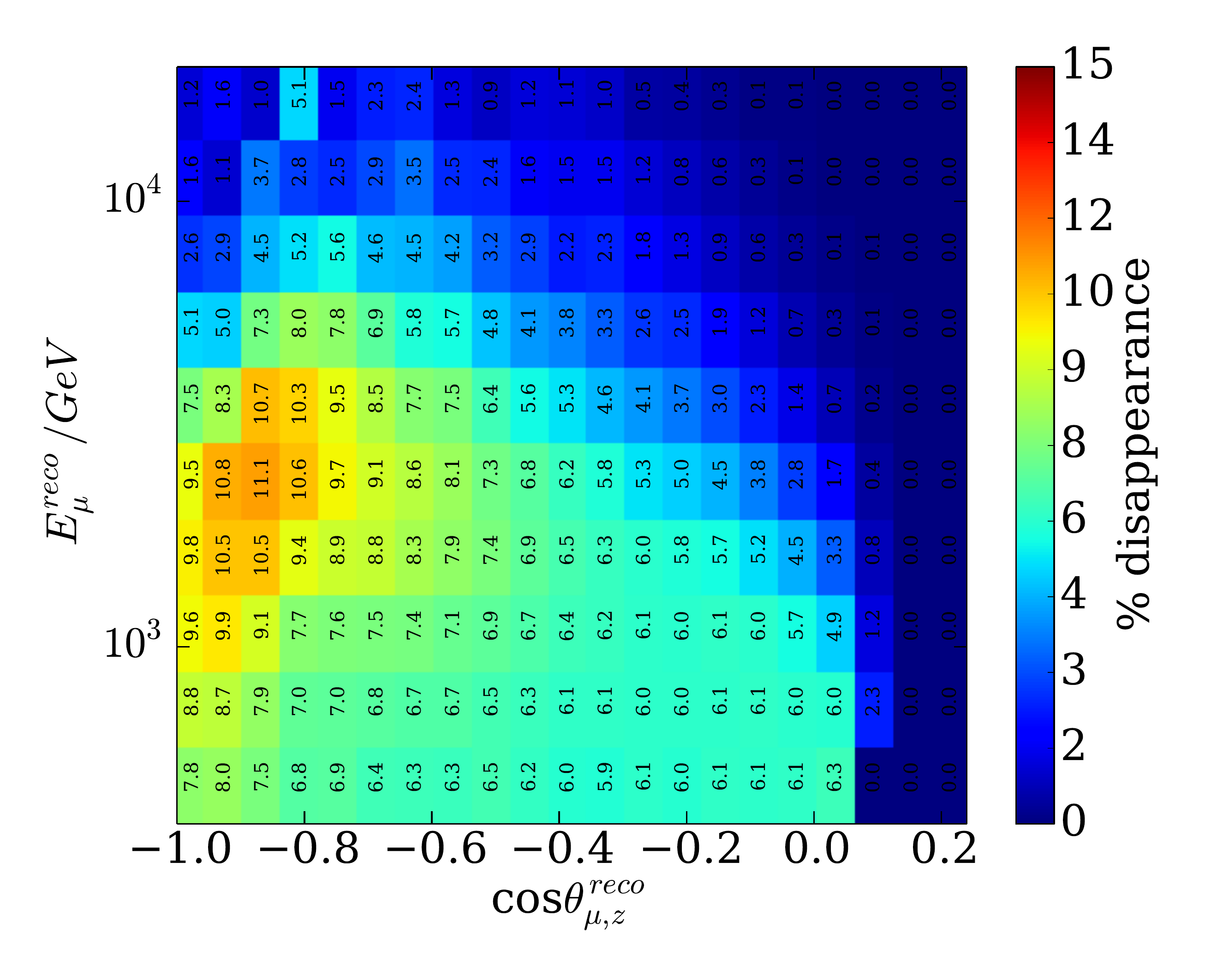}\\
\caption{Top and center: change in the spectrum due to propagation effects for muon neutrinos and antineutrinos at the 3+1 global best fit point. Bottom: The predicted event rate reduction (in percent) vs. reconstructed muon energy and zenith angle for this model.
\label{fig:Oscillograms}}
\end{figure}

\section{\label{sec:IceCube}Atmospheric Neutrinos in IceCube}

Having crossed the Earth, a small fraction of up-going atmospheric neutrinos undergo charged current interactions in either bedrock or ice, creating muons that traverse the instrumented ice of IceCube. These produce secondary particles that add Cherenkov light, which can be detected by the Digital Optical Modules (DOMs) \cite{Hanson:2006bk,Abbasi:2008aa,Abbasi:2010vc} of the IceCube array.  The full detector contains 5160 DOMs on 86 strings arranged with string-to-string spacing of approximately 125 m and typical vertical DOM separation of 17 m.  

The analysis detailed in this paper, referred to as IC86, uses data from the full 86-string detector configuration taken during 2011-2012, with up-going neutrinos selected according to the procedure developed in \cite{Weaver:thesis,Aartsen:2015rwa}. The sample contains 20,145 well-reconstructed muons detected over a live time of 343.7 days. A total of 99.9\% of the detected events in the data sample are expected to be neutrino-induced muon events from the decays of atmospheric pions and kaons. The flux contribution from charmed meson decays was found to be negligible \cite{Aartsen:2014muf,Aartsen:2015rwa}, as was the contamination of up-going astrophysical neutrinos with the spectrum and rate measured by IceCube \cite{Aartsen:2015rwa}. A complementary analysis, referred to as IC59 and discussed later, was performed using a sample of 21,857 events observed in 348.1 days of data taken with an earlier 59-string configuration of the detector from 2009-2010 \cite{Aartsen:2013eka}.

Since muon production is very forward at these energies, the muon preserves the original neutrino direction with a median opening angle following 0.7 degrees$\times$(E$_{\nu}$/TeV)$^{-0.7}$ \cite{Ahrens:2003ix}.
The muon zenith angle can be reconstructed geometrically with a resolution of $\sigma_{\mathrm{cos}(\theta_{z})}$ varying between 0.005 and 0.015 depending on the angle. Because of energy sharing in production and radiative losses outside the detector, the detected muon energy is smeared downward from the original neutrino value. Muon energy is reconstructed based on the stochastic light emission profile along the track \cite{Aartsen:2013vja,Aartsen:2015rwa} with a resolution of $\sigma_{\mathrm{log}_{10} (E_\mu / GeV)}\sim0.5$. 

To search for shape effects \cite{Esmaili:2013vza,Esmaili:2012nz,Esmaili:2013cja,Barger:2011rc}, including the MSW and parametric resonances, the analyses compare the predicted observable muon spectrum for a given incident neutrino flux and oscillation hypothesis with data. Flavor evolution in the active and sterile neutrino system can be calculated by numerical solution of a master equation \cite{GonzalezGarcia:2005xw,Delgado:2014kpa}.  For IC86, this calculation is performed using the $\nu$-\texttt{SQuIDs} software package \cite{squids,nusquids}, while the IC59 analysis approximates the oscillation probability by solving a Schr\"{o}dinger-like equation using the NuCraft package~\cite{Wallraff:2014qka}. This approximation is accurate to better than 10\% below $\Delta m^2 \approx 5~{\rm eV}^2$, where Earth-absorption effects can be neglected. Fig.~\ref{fig:Oscillograms} (top and center) shows the  $\nu_\mu$ and $\bar\nu_\mu$ oscillation probability vs. true energy and zenith angle, calculated at the best-fit point from \cite{Conrad:2012qt}. Since IceCube has no sign-selection capability, the reconstructed samples contain both $\mu^+$ and $\mu^-$ events.  For illustration, Fig. \ref{fig:Oscillograms} (bottom) shows the predicted depletion of  events for the global 3+1 best fit point in the distribution of reconstructed variables from the IC86 analysis;  in this case the large depletion is dominated by the parametric resonance.

\section{\label{sec:Systematics}Data Analysis and Systematic Uncertainties}

To search for sterile neutrino oscillations we calculate the negative of a binned Poissonian log-likelihood (LLH) for the data given each sterile neutrino hypothesis on a fine grid in the the $\left[ \mathrm {log}(\Delta m^2), \mathrm{log}(\mathrm{sin}^22\theta_{24}) \right]$ hypothesis space. In the IC86 analysis, the data are histogrammed on a grid with 10 bins in energy ranging from 400~GeV to 20~TeV, and 21 linearly spaced bins starting at $\mathrm{cos}(\theta) = 0.24$ with a spacing of 0.06. The bins were chosen a priori guided by experimental resolution, scale of the disappearance signatures and accumulated MC simulation statistics. The LLH values are compared to the minimum in the space to produce unified confidence intervals \cite{Feldman:1997qc}.
Systematic uncertainties are treated by introducing both continuous and discrete nuisance parameters, which are fitted at each hypothesis point. The list of systematic uncertainties considered is given in Table \ref{tbl:systematics} and discussed below.  More information can be found in \cite{BenThesis} and \cite{CarlosThesis}.

\begin{table}[h]
\begin{tabular}{ l  c  c } 
\hline
\multicolumn{3}{c}{\it Atmospheric flux} \\ \hline \hline
$\nu$ flux template & discrete (7) & \\
$\nu$ / $\overline{\nu}$ ratio & continuous & 0.025\\ 
$\pi$ / K ratio & continuous & 0.1\\
Normalization & continuous & none$^1$ \\ 
Cosmic ray spectral index  & continuous & 0.05 \\
Atmospheric temperature  & continuous & model tuned\\
\hline \hline
\multicolumn{3}{c}{\it Detector and ice model} \\ \hline \hline
DOM efficiency &  continuous &  \\
Ice properties &  discrete (4) & \\ 
Hole ice effect on angular response &  discrete (2)\\ \hline
\multicolumn{3}{c}{\it Neutrino propagation and interaction} \\ \hline \hline
DIS cross section &  discrete (6) & \\
Earth density & discrete (9) & \\
\hline \hline
\end{tabular}
\caption{List of systematic uncertainties considered in the analysis. The numbers in parentheses show the number of discrete variants used. Full descriptions are given in the text. The third column indicates the gaussian width of a prior if introduced for the parameter in the analysis (see \cite{BenThesis}   for details). $^1$A prior of 40\% was applied to the Normalization parameter in the rate+shape analysis described below.}
\label{tbl:systematics}
\end{table}

\subsection{Atmospheric neutrino flux uncertainties}

The atmospheric flux in the energies relevant to this analysis is dominated by the neutrinos that originate from pion and kaon decays in cosmic ray showers. This prompts us to parametrize the atmospheric flux as
\begin{equation}
\phi_{\rm atm}(\cos\theta) = N_0  \mathcal{F}(\delta)\bigg(  \phi_\pi  + R_{\pi/K} \phi_K \bigg) \left(\frac{E_\nu}{E_0}\right)^{- \Delta \gamma}
\label{eq:atmospheric-parametrization}
\end{equation}
\noindent (and similarly for antineutrinos, with a relative flux normalization uncertainty). The free nuisance parameters are the overall flux normalization $N_0$, the correction to the ratio of kaon- to pion-induced fluxes $R_{K/\pi}$ and the spectral index correction $\Delta \gamma$. The $\phi_\pi$ and $\phi_K$ are the spectrum of atmospheric neutrinos originating from $\pi$ and $K$ decays, respectively. Furthermore, $\Delta \gamma$ allows us to take into account uncertainties in the spectral index of the flux. The term $E_0$ is a pivot point near the median of the energy distribution which renders the $\Delta \gamma$ correction approximately normalization-conserving.

Here, seven $\phi_k$ and $\phi_\pi$ variants are used to encapsulate additional hadronic model uncertainty and the primary cosmic ray model uncertainties. Atmospheric density uncertainties are a subleading effect. We thus parametrize it as a linear function, $\mathcal{F}(\delta)$, which is obtained by fitting fluxes calculated with different atmospheric profiles generated within constraints imposed by temperature data from the AIRS satellite \cite{AIRS}. 

The central flux prediction for the analysis is the HKKM model with H3a knee correction \cite{Sanuki:2006yd,Honda:2006qj,Gaisser:2013bla}.  Additional flux variants are calculated using the analytic air shower evolution code of \cite{Fedynitch:2015zma,MCeq,Collins2015}. The cosmic spectrum variants considered are the Gaisser-Hillas \cite{Gaisser:2013bla}, Zatsepin-Sokolskaya \cite{Zatsepin:2006ci}, and Poly-gonato models \cite{Hoerandel:2002yg}.  The hadronic models considered are QGSJET-II-4 \cite{Ostapchenko:2010vb} and SIBYILL2.3 \cite{Riehn:2015oba}.  For each combination of hadronic and primary model, fluxes calculated in various atmospheric density profiles are used to derive the  $\mathcal{F}(\delta)$ parameterization.

\subsection{Neutrino propagation and interaction uncertainties}

Two sets of neutrino propagation uncertainties are treated in the search. Neutrino oscillation and absorption effects both depend on the Earth density profile along the neutrino trajectory, which is parameterized by the PREM model \cite{ref:PREM}. Uncertainties in the Earth composition and density are accounted for by creating perturbations of the PREM and re-propagating the neutrino flux. The PREM variants are constructed under the constraints that the Earth mass and moment of inertia are preserved, that the density gradient is always negative in the core and mantle regions, and that the local perturbation is never more than 10\%. The effects of Earth model uncertainty on the final propagated neutrino spectrum are incorporated by minimizing over 9 discrete perturbed models.  

A further propagation uncertainty is the neutrino charged-current cross-section that, at these energies, is dominated by deep inelastic scattering (DIS). The uncertainty in the cross-sections arises from parton distribution function (PDF) uncertainties. A parametrization of the cross-section uncertainty uses calculations \cite{CooperSarkar:2011pa} (see also \cite{ref:CarlosXS}) based on three different PDF sets: HERAPDF \cite{Aaron:2009aa}, CT10 \cite{Gao:2013xoa} and NNPDF \cite{,Nocera:2014gqa}. In each case, simulated neutrino interactions are re-weighted using true neutrino energy and inelasticity given calculated doubly-differential cross sections, and the analysis fit is run using the weighted sample.

\subsection{Detector and ice uncertainties}

The absolute optical module photon collection efficiency, $\epsilon$, has been measured in the laboratory \cite{Abbasi:2010vc}. However, shadowing by the DOM cable and unknown local optical conditions after deployment introduce an uncertainty in the optical efficiency {\it in situ}, leading to uncertainty in the detected energy and angular event distribution.  Here  $\epsilon$ is treated as a continuous nuisance parameter and re-weighting techniques are used to correct Monte Carlo distributions to arbitrary values. We follow the method developed in \cite{Weaver:thesis,Aartsen:2015rwa}, implementing a penalized spline \cite{Whitehorn:2013nh} fitted to Monte Carlo datasets generated at various DOM efficiency values. Variability of the optical efficiency induces changes in the detector energy scale.  In practice, the best fit value is tightly constrained by the position of the energy peak in the final sample.

The IceCube ice model applied in this analysis has nearly a thousand free parameters that are minimized in an iterative fit procedure using light-emitting-diode (LED) flasher data \cite{Aartsen:2013rt}. The model implements vertically varying absorption and scattering coefficients across tilted isochronal ice layers. The fit procedure yields a systematic and statistical uncertainty on the optical scattering and absorption coefficients in the ice, as well as a larger uncertainty on the amount of light deposited by the LED flashers. This larger uncertainty was later reduced by introducing azimuthal anisotropy in the scattering length into the ice model, which may result from dust grain shear due to glacial flow \cite{Aartsen:2013ola}. We use the model described in \cite{Aartsen:2013rt} as the central ice model, and then use the model with anisotropy \cite{Aartsen:2013ola} as an alternative to assess the impact of this effect. We also incorporate models with 10\% variations in the optical absorption and scattering coefficients to account for the uncertainty on those parameters. A full Monte Carlo sample is created for each model variation.

The ice column immediately surrounding the DOMs has different optical properties than the bulk ice due to dissolved gases that are trapped during the refreezing process following DOM deployment. This introduces additional scattering near the DOM and has a nontrivial effect on its angular response \cite{Aartsen:2013rt}. To quantify this effect on the final event distribution, a comparison is made between the extreme case of the DOM assumed to have its laboratory-derived angular response vs.\ the nominal hole ice model as discrete ice model variants.

\section{\label{sec:Results}Results}

\begin{figure}[t]
\includegraphics[width=0.90\columnwidth]{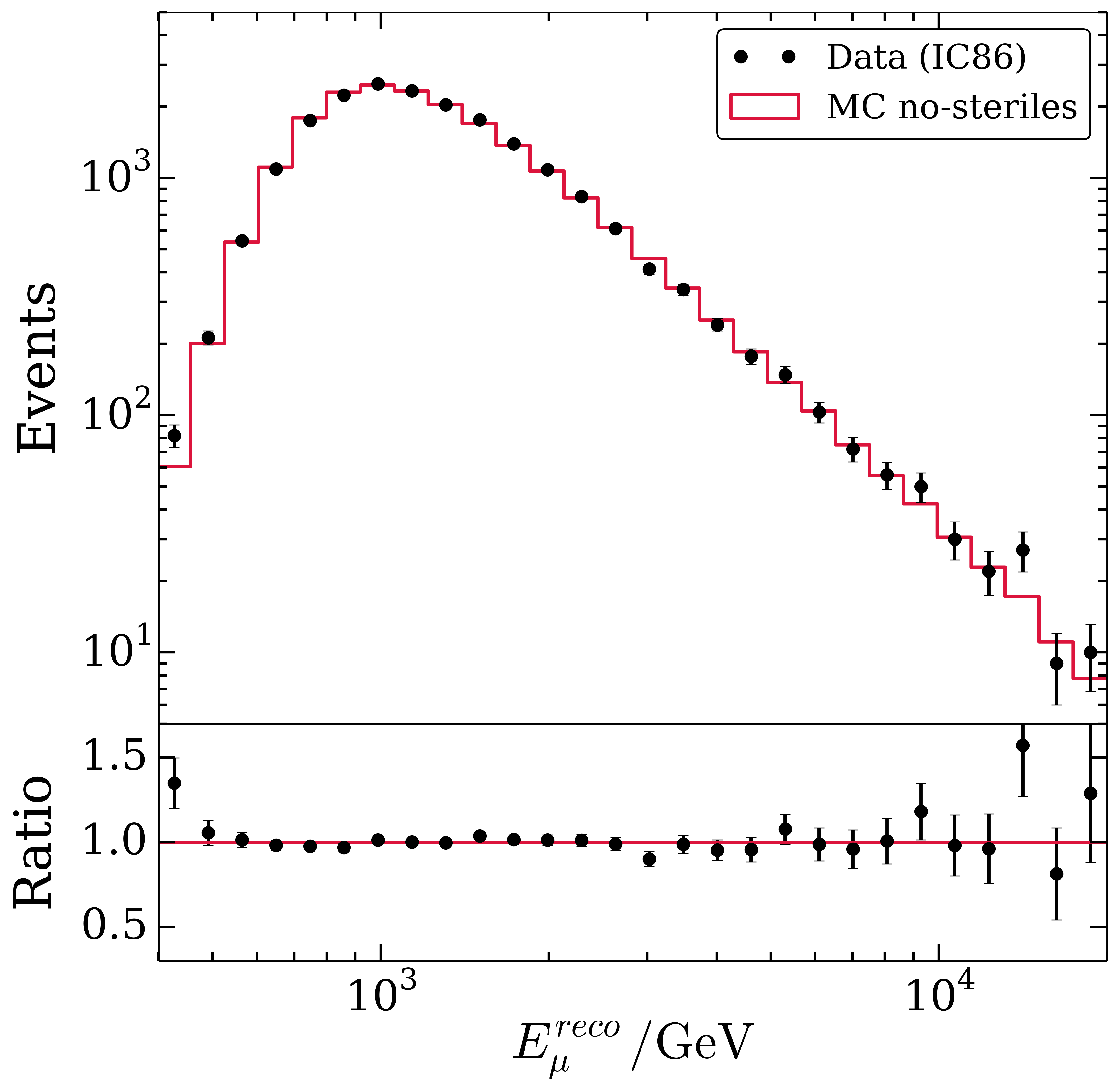} \\
\caption{\label{fig:PreUnblinding}Reconstructed energy distribution in data and Monte Carlo for the no-steriles hypothesis in the analysis.}
\end{figure}

\begin{figure}[t]
\includegraphics[width=0.99\columnwidth]{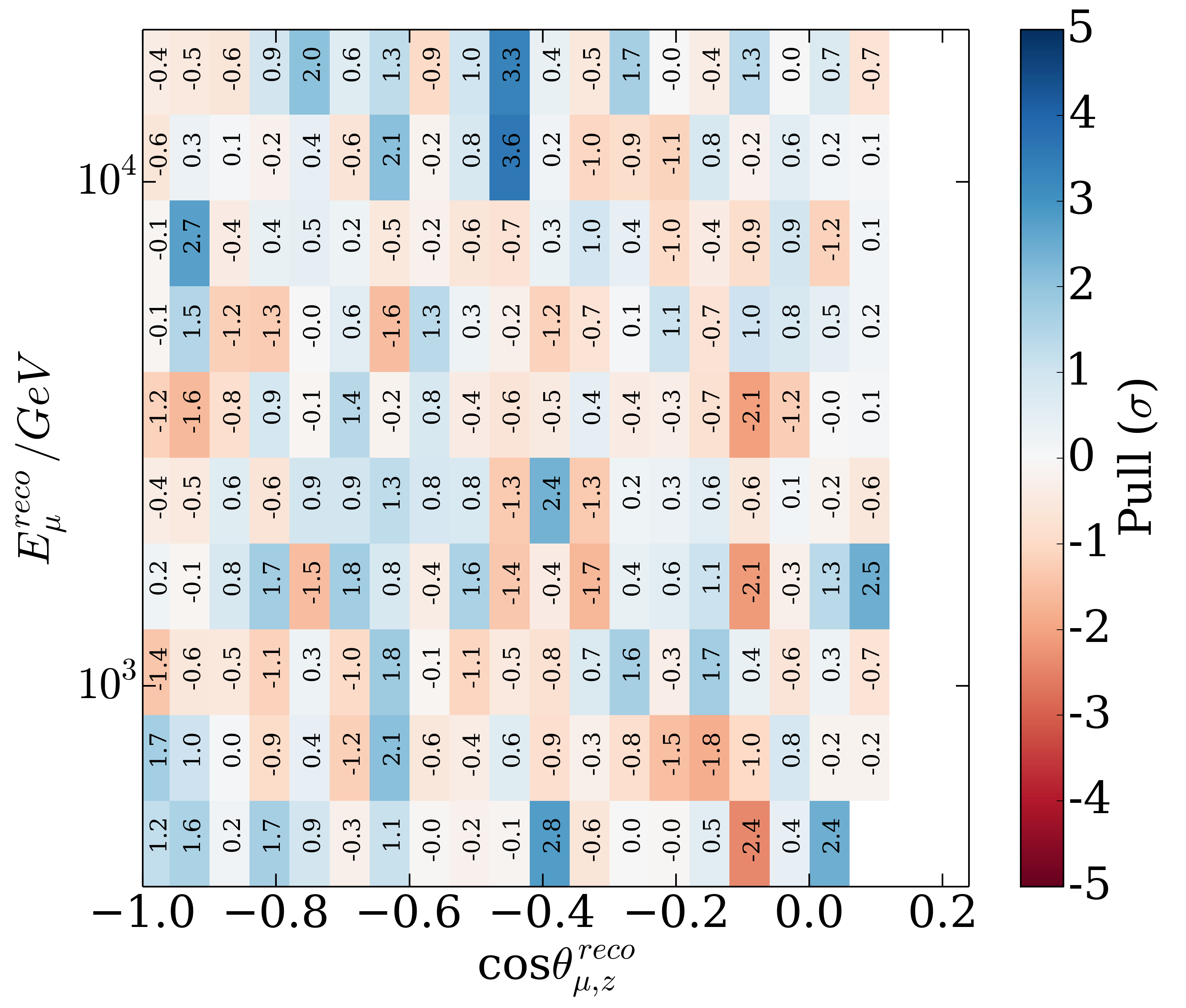}\\
\caption{\label{fig:Pulls}The statistical-only pulls (shape+rate analysis) per reconstructed energy and zenith angle bin at the best nuisance parameter fit point for the no-sterile hypothesis. The shown empty bins are those that were evaluated in the analysis but had no data events remaining following cuts.}
\end{figure}

The analysis detailed here was developed with 90\% of the data sample held blind, and unblinding was a multi-step process. The agreement of Monte Carlo (MC) simulations based on the no-steriles hypothesis (corresponding to more than 360 years of simulated data) with data was evaluated using one-dimensional energy and zenith angle distributions, which would wash out the resonance signature of sterile neutrinos (Fig.~\ref{fig:PreUnblinding}). Good data-MC consistency was observed and no nuisance parameter was found to have a significant pull outside of its prior. Other comparisons, insensitive to the sterile neutrino signature, were made by examining subsets of the data split by reconstructed azimuthal track angle, and by event center-of-gravity. No significant data-Monte Carlo disagreements were observed. The full event distribution in the two-dimensional analysis space, and the pulls-per-bin from the null hypothesis (Fig.~\ref{fig:Pulls}) were then examined. Event-by-event reconstructed data and Monte Carlo can be found in \cite{datarelease}.

The LLH value for the data given each sterile neutrino hypothesis was calculated.  No evidence for sterile neutrinos was observed. The best fit of the blind, shape-only analysis is at $\Delta m^2$ = 10 eV$^2$ and $\mathrm{sin}^2 2\theta_{24}$=0.56 with a log likelihood difference from the no-steriles hypothesis of $\Delta \rm{LLH}$=1.91, corresponding to a p-value of 15\%.  Since the fit does not constrain flux normalization, LLH minima at $\Delta m^2 \gtrsim 5~{\rm eV^2}$ are highly degenerate with the no-sterile hypothesis.  This is because the oscillation effect becomes a fast vacuum-like oscillation smeared out by the energy resolution of the detector, and thus changes the normalization but has no effect on shape. 

\begin{figure}[h!]
\includegraphics[width=0.99\columnwidth]{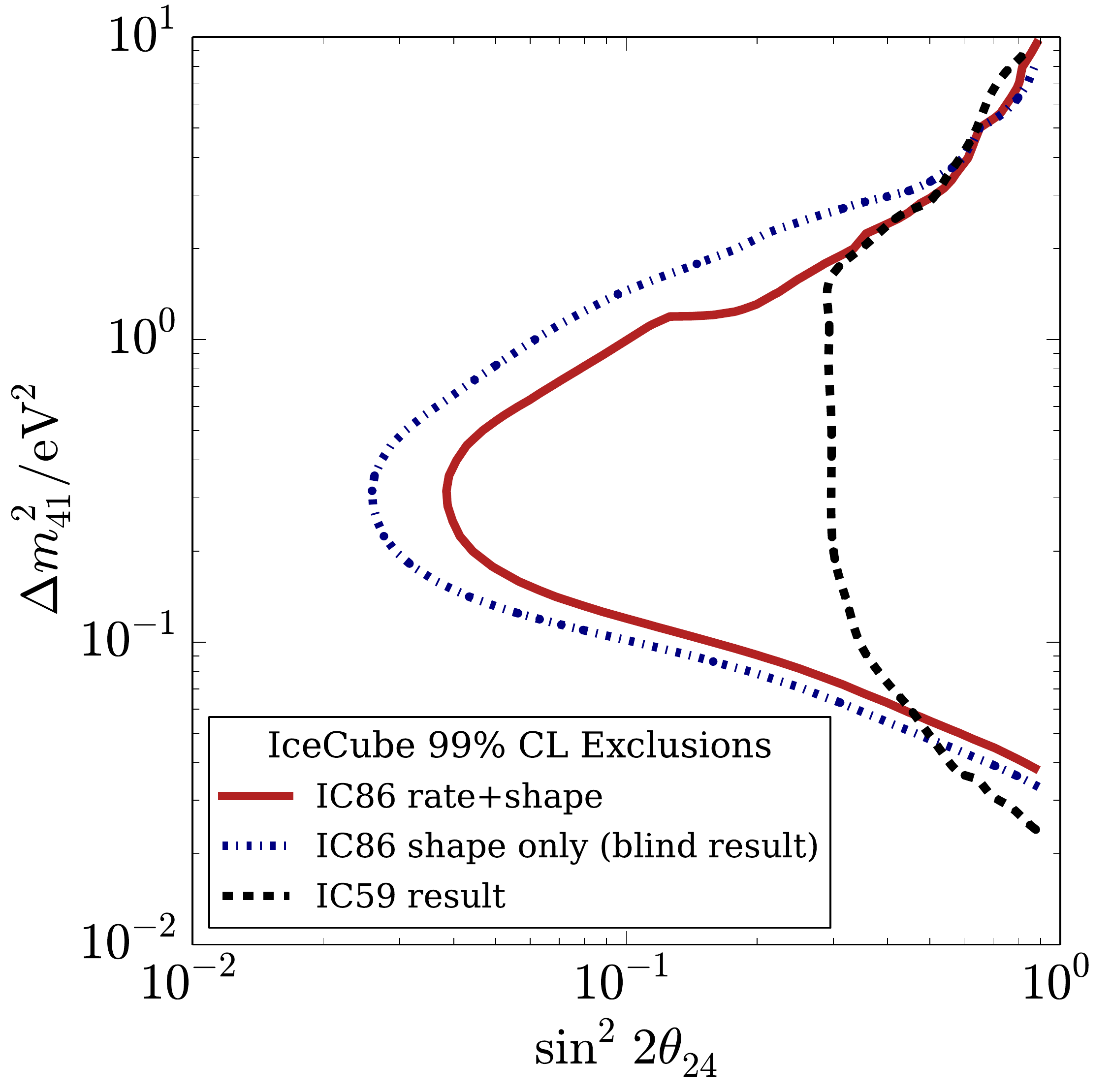}
\caption{\label{fig:IceCubeResults}Results from IceCube sterile neutrino searches (regions to the right of the contours are excluded). The dot-dashed blue line shows the result of the original analysis based on shape alone, while the solid red line shows the final result with a normalization prior included to prevent degeneracies between the no-steriles hypothesis and sterile neutrinos with masses outside the range of sensitivity.  The dashed black line is the exclusion range derived from an independent analysis of data from the 59-string IceCube configuration.   
}
\end{figure}

Post-unblinding tests highlighted two undesirable features of the shape-only analysis, both deriving from the degeneracy between high-$\Delta m^2$, fast oscillation hypotheses and changes in the flux normalization.  First, because the high-$\Delta m^2$ space is not penalized by any prior, a log likelihood minimum in this region may not be uniquely defined under extensions of the search space.  In some cases, slightly stronger exclusion limits can be found by increasing the search space to higher mass.  Second, the degeneracy between normalization and mixing can lead to 
unphysical values for the normalization that compensate for the sterile neutrino oscillation effect. To avoid these ambiguities, an extension of the analysis (denoted rate+shape) was developed to constrain the neutrino flux normalization using a prior with 40\% uncertainty in the likelihood function, based on \cite{Fedynitch:2012fs,Honda:2006qj}. This results in a weakened exclusion relative to the blind analysis proposal. However, since it is more robust, we consider it our primary result. For the rate+shape analysis, the best fit is at $\Delta m^2=10$~eV$^2$ and sin$^2 2\theta_{24}$=0.50, with a log likelihood difference from the no-steriles hypothesis of $\Delta$LLH=0.75, corresponding to a p-value of 47\%. This minimum is unique under extension
of  the  analysis  space  to  higher  masses,  since  the  large $\Delta m^2$ region  is  no  longer  degenerate  with  the  no-sterile hypothesis. This was checked over an extended parameter space up to $\Delta m^2$=100 ${\rm eV}^2$. The confidence interval for the shape-only and the rate+shape analyses are shown in Fig.~\ref{fig:IceCubeResults}.

A number of checks of the rate+shape analysis result were made (see \cite{BenThesis}). The exclusion is found to be robust under tightening or loosening the nuisance parameter priors by a factor of two. Different strengths of the normalization constraint were tested, and the result was found to be relatively insensitive to values between 30\% and 50\%.The pulls on each continuous nuisance parameter were evaluated at all points in the LLH space and found to behave as expected. The contour was redrawn for each discrete nuisance variant and found to have good stability. The Wilks confidence intervals \cite{Agashe:2014kda} were validated using Feldman-Cousins ensembles along the contour \cite{Feldman:1997qc} and found to be accurate frequentist confidence intervals.

An independent search was conducted using the 59-string IceCube data \cite{Aartsen:2015cwa,MariusThesis}, introduced previously, that also finds no evidence of sterile neutrinos. The IC59 analysis, described in detail in \cite{Aartsen:2013eka}, used different treatments for the systematic uncertainties, for the fitting methods and employed independent Monte Carlo samples that were compared to data using unique weighting methods. In particular, the event selection used for this data set had higher efficiency for low-energy neutrinos, using a threshold at 320 GeV, extending the sensitivity of the analysis to smaller $\Delta m^2$. However, detailed {\it a posteriori} inspections revealed that a background contamination from cosmic ray induced muons, on the level of 0.3\% of the full sample, is largest in this region and could lead to an artificially strong exclusion limit. Furthermore, the energy reconstruction algorithm used in both analyses, which measures the level of bremsstrahlung and other stochastic light emission along the muon track, is vulnerable to subtle detector modeling issues and suffers degraded energy resolution in the low-energy region where most muons are minimum-ionizing tracks and a large fraction either start or stop within the detector. It was therefore decided to exclude these events to avoid biasing the resulting exclusion regions. As a result of this {\it a posteriori} change, the IC59 analysis retains a comparable range of sensitivity in $\Delta m^2$ but the reach in $\rm{sin}^2$$\theta_{24}$ is strongly reduced (see Fig.~\ref{fig:IceCubeResults}). However, we still present this result as it independently confirms the result presented here.

\section{\label{sec:Conclusions}Discussion and Conclusion}

Resonant oscillations due to matter effects would produce distinctive signatures of sterile neutrinos in the large set of high energy atmospheric neutrino data recorded by the IceCube Neutrino Observatory.  The IceCube collaboration has performed searches for sterile neutrinos with $\Delta m^2$ between $0.1$ and $10 ~{\rm eV^2}$. We have assumed a minimal set of flavor mixing parameters in which only $\theta_{24}$ is non-zero. 

A nonzero value for $\theta_{34}$ would change the shape of the MSW resonance while increasing the total size of the disappearance signal \cite{Esmaili:2013vza}. As discussed in \cite{Lindner:2015iaa}, among the allowed values of $\theta_{34}$ \cite{Adamson:2011ku}, the model with $\theta_{34}$=0 presented here leads to the most conservative exclusion in $\theta_{24}$. The angle $\theta_{14}$ is tightly constrained by electron neutrino disappearance measurements \cite{Kopp:2013vaa}, and nonzero values of $\theta_{14}$ within the allowed range do not strongly affect our result.

\begin{figure}[tbp!]
\includegraphics[width=0.99\columnwidth]{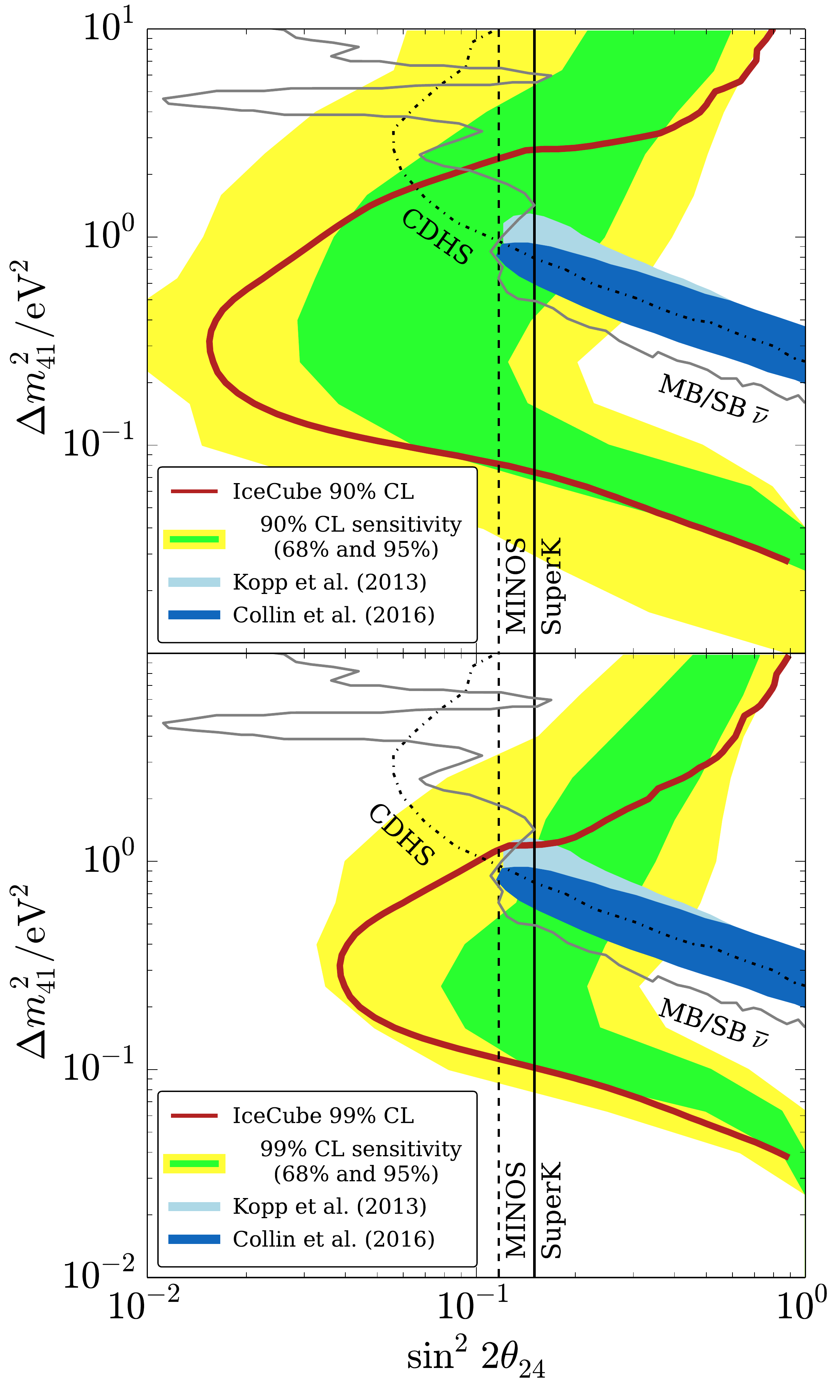}\\
\caption{\label{fig:RateShapeExcl}Results from the IceCube search. (Top) The 90\% (orange solid line) CL contour is shown with bands containing 68\% (green) and 95\% (yellow) of the 90\% contours in simulated pseudo-experiments, respectively. (Bottom)  The 99\% (red solid line) CL contour is shown with bands containing 68\% (green) and 95\% (yellow) of the 99\% contours in simulated pseudo-experiments, respectively. The contours and bands are overlaid on 90\% CL exclusions from previous experiments \cite{Abe:2014gda, Adamson:2011ku, Cheng:2012yy, Dydak:1983zq}, and the 99\% CL allowed region from global fits to appearance experiments including MiniBooNE and LSND, assuming $|U_{e4}|^2=0.023$ \cite{Kopp:2013vaa} and $|U_{e4}|^2=0.027$ \cite{Conrad:2012qt} respectively.}
\end{figure}

Figure~\ref{fig:RateShapeExcl} shows the current IceCube results at 90\% and 99\% confidence levels, with predicted sensitivities, compared with 90\% confidence level exclusions from previous disappearance searches \cite{Abe:2014gda, Adamson:2011ku, Cheng:2012yy, Dydak:1983zq}. Our exclusion contour is essentially contained within the expected +/- 95\% range around the projected sensitivity derived from simulated experiments, assuming a no-steriles hypothesis. In any single realization of the experiment, deviations from the mean sensitivity are expected due to statistical fluctuations in the data and, to a considerably lesser extent, in the Monte Carlo data sets. Also shown is the 99\% allowed region from a fit to the short baseline appearance experiments, including LSND and MiniBooNE, from \cite{Kopp:2013vaa, Esmaili:2013vza, Conrad:2012qt}, projected with $|U_{e4}|^2$ fixed to its world best fit value according to global fit analyses \cite{Kopp:2013vaa,Conrad:2012qt,Collin:2016rao}. This region is excluded at approximately the 99\% confidence level, further increasing tension with the short baseline anomalies, and removing much of the remaining parameter space of the 3+1 model. 
We note that the methods developed for the IC59 and IC86 analyses are being applied to additional data sets, including several years of data already recorded by IceCube, from which we anticipate improvements in IceCube’s sterile neutrino sensitivity.

\begin{acknowledgments}

We acknowledge the support from the following agencies:
U.S. National Science Foundation-Office of Polar Programs,
U.S. National Science Foundation-Physics Division,
University of Wisconsin Alumni Research Foundation,
the Grid Laboratory Of Wisconsin (GLOW) grid infrastructure at the University of Wisconsin - Madison, the Open Science Grid (OSG) grid infrastructure;
U.S. Department of Energy, and National Energy Research Scientific Computing Center,
the Louisiana Optical Network Initiative (LONI) grid computing resources;
Natural Sciences and Engineering Research Council of Canada,
WestGrid and Compute/Calcul Canada;
Swedish Research Council,
Swedish Polar Research Secretariat,
Swedish National Infrastructure for Computing (SNIC),
and Knut and Alice Wallenberg Foundation, Sweden;
German Ministry for Education and Research (BMBF),
Deutsche Forschungsgemeinschaft (DFG),
Helmholtz Alliance for Astroparticle Physics (HAP),
Research Department of Plasmas with Complex Interactions (Bochum), Germany;
Fund for Scientific Research (FNRS-FWO),
FWO Odysseus programme,
Flanders Institute to encourage scientific and technological research in industry (IWT),
Belgian Federal Science Policy Office (Belspo);
University of Oxford, United Kingdom;
Marsden Fund, New Zealand;
Australian Research Council;
Japan Society for Promotion of Science (JSPS);
the Swiss National Science Foundation (SNSF), Switzerland;
National Research Foundation of Korea (NRF);
Villum Fonden, Danish National Research Foundation (DNRF), Denmark

\end{acknowledgments}

Note added:  Recently, an analysis using IceCube public data \cite{weaverdatarelease} was performed \cite{Liao:2016reh}.  Though this independent analysis has a limited treatment of systematics, it follows the technique described here and in refs. \cite{BenThesis, CarlosThesis}, and obtains comparable bounds. To allow for better reproduction of the result shown in this paper in the future,  we have put forward a data release that incorporates detector systematics \cite{datarelease}.

\bibliography{apssamp}

\end{document}